\documentclass[conference]{IEEEtran}
\usepackage{cite}

\ifCLASSINFOpdf
\else
\fi
\usepackage{amsmath}
\usepackage{algorithmic}

\usepackage{array}
 \usepackage{dblfloatfix}

\usepackage{url}
\usepackage[bookmarks=false,draft]{hyperref}
\usepackage[utf8x]{inputenc}
\usepackage[T1]{fontenc}
\usepackage{lmodern}
\usepackage{fourier, erewhon} \usepackage{multirow,bigdelim}
\usepackage{mathtools, blkarray}
\usepackage{xfrac}

\usepackage{commath}
\usepackage{graphicx}
\usepackage[margin=1in]{geometry}
\usepackage{multirow,booktabs}
\usepackage{epstopdf}
\DeclareGraphicsExtensions{.eps}
\usepackage{amsfonts,amssymb,amsthm,epsfig,url,array}
\usepackage[space]{grffile}
\theoremstyle{plain}
\newtheorem{theorem}{Theorem}
\usepackage{chngcntr}
\usepackage{apptools}
\AtAppendix{\counterwithin{lemma}{section}}
\newtheorem{lemma}{Lemma}
\usepackage{float}
\usepackage{textcomp}
\usepackage{gensymb}
\usepackage{tabularx,ragged2e,booktabs,caption}
\usepackage{amssymb}
\usepackage{lmodern}
\usepackage{fixltx2e}
\usepackage[font=scriptsize]{caption}
\usepackage{hyperref}
\newcolumntype{C}[1]{>{\Centering}m{#1}}

\usepackage{etoolbox}

\begin{document}
%
\title{A novel CS Beamformer root-MUSIC algorithm and its subspace deviation analysis}
\author{\IEEEauthorblockN{$^\dagger$Abhishek Aich, \textit{Student Member} $^\ddagger$P. Palanisamy, \textit{Member, IEEE}}
\IEEEauthorblockA{Department of Electronics and Communication Engineering, \\
National Institute of Technology, Tiruchirappalli-620 015, India \\
email: $^\dagger$abhishekaich.nitt@gmail.com, $^\ddagger$palan@nitt.edu}

}
\maketitle
\thispagestyle{empty}
\begin{abstract}
Subspace based techniques for direction of arrival (DOA) estimation need large amount of snapshots to detect source directions accurately. This poses a problem in the form of computational burden on practical applications. The introduction of compressive sensing (CS) to solve this issue has become a norm in the last decade. In this paper, a novel CS beamformer root-MUSIC algorithm is presented with a revised optimal measurement matrix bound. With regards to this algorithm, the effect of signal subspace deviation under low snapshot scenario (e.g. target tracking) is analysed. The CS beamformer greatly reduces computational complexity without affecting resolution of the algorithm, works on par with root-MUSIC under low snapshot scenario and also, gives an option of non-uniform linear array sensors unlike the case of root-MUSIC algorithm. The effectiveness of the algorithm is demonstrated with simulations under various scenarios.\\*

\end{abstract}


%
\IEEEpeerreviewmaketitle

\section{Introduction}
Direction of arrival (DOA) estimation is needed in major applications like dynamic tracking and surveillance in radar systems \cite{naidu}. In such application areas, data acquisition using Nyquist rate is found to be sub-optimal because for proper detection using such an data rate, a large number of snapshots are required. This makes the process too costly; ending up with too many samples with separate requirement of building instruments capable of acquiring such data at the required rate. Hence, though there are  many advances being made to handle such large data, especially for memory power, it is intriguing challenge to get better results with less complexity. One of the techniques being used to overcome such an challenge is the introduction of compressive sensing (CS)\cite{candes} to DOA estimation.

Compressive sensing represents a paradigm of undersampling a given signal and then retrieving it by finding the most sparsest representation of the signal, with pre-agreed error. The CS beamformer was proposed in \cite{ying} which modifies the DOA data model to CS-DOA data model and then solves it with traditional estimation algorithms \cite{naidu, schmidt}. This leads to reduced computational complexity of these algorithms. In \cite{aich}, it is shown that a more strict bound can be imposed on the measurement matrix used in the CS-DOA data model, contrary to the traditional CS bound of \textit{O}(\textit{s}[\textit{ln}\textit{N}]), on a \textit{s}-sparse signal  $\textbf{x} \in \displaystyle\mathbb{C}^{\textit{N}\times1}$. In this paper, a novel CS beamformer root-MUSIC algorithm is proposed with this similar bound which shows almost comparable performance with the root-MUSIC algorithm. The subspace deviation is also analysed w.r.t. the new data model for this proposed algorithm. Note that throughout this paper, $ [\cdot]^{T}$ represents transpose of a matrix, $[\cdot]^{H}$ represents \textit{Hermitian} meaning conjugate transpose of a complex matrix, $\text{Tr}[\cdot]^{H}$ represents trace of the matrix and $\textbf{I}_K$ represents identity matrix of order \textit{K}.

This paper is organized as follows. Section II gives the proposed CS beamformer root-MUSIC algorithm. In Section III, the subspace deviation due to low snapshot in case of CS beamformer is derived. Simulation results and analysis is provided in Section IV, and conclusion in Section V.


 
\section{Proposed Algorithm}
\subsection{Brief review of Compressive Sensing}
Compressive sensing was introduced formly by Candès and Wakin in \cite{candes}. It states that to reconstruct a \textit{s}-sparse signal \textbf{x}  $\in \displaystyle\mathbb{C}^{\textit{N$\times${1}}}$ (\(\textit{s} < \textit{N}\)) from  a measured signal \textbf{y}  $\in \displaystyle\mathbb{C}^{\textit{m$\times${1}}}$, only \(\textit{m} \geq \textit{s} [ln(\textit{N})]\) number of measurements are needed. Even if \textbf{x} is not sparse in the current form, a sparse representation of the signal can be used in its place (as in the case of DOA model). 
This is via the transformation
\begin{equation}
\textbf{y} = \textbf{$\Phi$}\textbf{x}
\end{equation}
\noindent where $\Phi \in \displaystyle\mathbb{C}^{\textit{m$\times${N}}}$ is called the measurement matrix. Generally, retrieving \textbf{x} or its sparse form represents a convex optimization problem which can be solved with suitable algorithms. For reader's clarity, it is pointed out that (1) is given for one snapshot. Further discussion in the paper is for more than one snapshot. 

\subsection{CS beamformer root-MUSIC Algorithm}
In this section, the proposed CS beamformer root-MUSIC algorithm is discussed with the bounds on measurement matrix derived in \cite{aich}. 
Let a uniform linear array (ULA) be composed of
\textit{N} isotropic sensors. Let \textit{M} be the number of narrow-band signals scanned by the ULA from distinct directions denoted as \(\theta=(\theta_1, \theta_1,..., \theta_{\textit{M}}) \). Then, the output sample of the $\textit{n}^{th}$  sensor mixed with additive white noise, at the $\textit{t}^{th}$ instant is given as
\begin{equation}
\textit{x}_{n}(\textit{t}) = \sum_{\textit{i}=1}^{\textit{M}}\textit{a}_{n}(\theta_i)\textit{s}_i(\textit{t}) + \textit{w}_{n}(\textit{t});\quad \textit{t} = 1,2,.....,\textit{T}
\end{equation}
\textit{T} denotes the number of snapshots, $\textit{s}_i(\textit{t})$ is the $\textit{t}^{th}$ sample of the $\textit{i}^{th}$ signal, $ \textit{w}_n(\textit{t}) $ is the  $\textit{t}^{th}$ noise sample at the $\textit{n}^{th}$ sensor, \(\textit{a}_{n}(\theta_i) = \textit{e}^{-j(n-1)\gamma_{i}}\) where $\gamma_{i}$ = $\frac{{2}\pi{d}}{\lambda}$ sin($\theta_{i}$), \textit{d} and $\lambda$ are inter-element spacing and wavelength, respectively. Writing (2) in matrix notation, we have
\begin{equation}
\textbf{x}(t) = \textbf{A}(\textbf{$\theta$})\textbf{s}(t) + \textbf{w}(t);\quad \textit{t} = 1,2,.....,\textit{T}
\end{equation}
where, \\$\textbf{x}(t)=[\textit{x}_{1}(\textit{t}), \textit{x}_{2}(\textit{t}),\hdots ,\textit{x}_{N}(\textit{t})]^T$\\$\textbf{A}(\textbf{$\theta$})=[\textbf{a}(\theta_{1}),\textbf{a}(\theta_{2}),\hdots,\textbf{a}(\theta_{M})]$\\$\textbf{s}(t)=[\textit{s}_{1}(\textit{t}), \textit{s}_{2}(\textit{t}),\hdots ,\textit{s}_{M}(\textit{t})]^T$\\ $\textbf{w}(t)=[\textit{w}_{1}(\textit{t}), \textit{w}_{2}(\textit{t}),\hdots ,\textit{w}_{N}(\textit{t})]^T $\\
$\textbf{a}(\theta_i)=[\textit{a}_1(\theta_i),\textit{a}_2(\theta_i),\hdots,\textit{a}_N(\theta_i)]^T$;\quad \textit{i} = 1,2,.....,\textit{M}\\

\noindent Here, $\textbf{x} \in \displaystyle\mathbb{C}^{\textit{N$\times$T}}$,  $\textbf{A} \in \displaystyle\mathbb{C}^{\textit{N$\times$M}}$,  $\textbf{s} \in \displaystyle\mathbb{C}^{\textit{M$\times$T}}$ and  $\textbf{w} \in \displaystyle\mathbb{C}^{\textit{N$\times$T}}$. It is assumed that sources are uncorrelated and the noise is independent from source, with zero-mean and variance $\sigma_n^2$.
To modify the DOA model to CS-DOA model, use (3) in (1). We get 
\begin{equation}
\textbf{y} = \Phi\textbf{A}\textbf{s} + \Phi\textbf{w}
\end{equation}
Then auto-correlation matrix of \textbf{y} is given as
\begin{equation}
\textbf{R}_{yy} = \Phi\textbf{R}_{xx}\Phi^{H}
\end{equation}
with 
\begin{equation}
\textbf{R}_{xx} = \textit{E}[\textbf{xx}^H]
\end{equation}
Though $\textbf{R}_{xx} \in \displaystyle\mathbb{C}^{\textit{N$\times$N}}$, we have $\textbf{R}_{yy} \in \displaystyle\mathbb{C}^{\textit{m$\times$m}}$ where \(m= M +1\). Further remarks on this value of \textit{m} is discussed in next subsection.

The eigenvalue decomposition of $\textbf{R}_{yy}$ is performed and after sorting in increasing order, first (\textit{m}-\textit{M}) eigenvectors, corresponding to first (\textit{m}-\textit{M}) eigenvalues, are arranged as columns of a matrix, say $\textbf{Q}_y \in \displaystyle\mathbb{C}^{\textit{M$\times$(m-M)}}$. $\textbf{Q}_y$ represents the noise subspace. Define \textit{z} $\triangleq \frac{{2}\pi{d}}{\lambda}$ sin($\theta$). Then, the steering vector given $\textbf{a}(\theta)$ is represented as
\begin{equation}
\textbf{a}(z) = [1, z^{-1}, ..., z^{-(M-1)}] 
\end{equation}
Let \(\textbf{b}(z)=\Phi\textbf{a}(z)\). Then, according to \cite{bara}, argument of the roots of the equation \begin{equation}
\textbf{b}^T(z^{-1})\textbf{Q}_y\textbf{Q}_y^H\textbf{b}(z)=0
\end{equation} 
gives the DOA estimates.

\subsection{Remarks on m = M+1}
The bound \textit{m} of the measurement matrix was found to be sub-optimal, \textit{i.e.} it can further be decreased below the conventional CS theory. A strict bound on \textit{m} was hence, proposed given by the following theorem (Lemma 1 in \cite{aich}):
\begin{theorem}
If the number of sources to locate is \textit{M} and the number of sensors is \textit{N} (\textit{N} > \textit{M}), then the number of compressed measurements \textit{m} of the \textit{M}-sparse signal can be selected from the bound
\begin{flalign*}
  m &\in ( \textit{M}, \textit{N})
\end{flalign*}
\end{theorem}
It can be directly concluded from the above theorem that the optimal value of \textit{m} for accurate detection of DOAs can be taken as \textit{M}+1. This theorem also agrees for the root-MUSIC algorithm which can be shown easily in the following.
\begin{proof}
Consider the dimensional bound of $\Phi$ to be  \textit{m} = M (number of sources). This implies $\textbf{R}_{yy} \in \displaystyle\mathbb{C}^{\textit{M$\times$M}}$. For \(\textbf{b}^T(z^{-1})\textbf{Q}_y\textbf{Q}_y^H\textbf{b}(z)\) to be zero, its roots have to be near the unit circle. This is only possible when atleast one noise eigenvector is present as the column in $\textbf{Q}_y$. But since, number of sources is equal to dimension of $\textbf{R}_{yy}$, there will be no noise eigenvector in $\textbf{Q}_y$. Therefore, (8) won't have any roots near the unit circle in \textit{z}-domain. Hence, \textit{m} should be atleast \textit{M} + 1.      
\end{proof}
\subsection{Remarks on computational complexity}
As seen above, the measurement matrix changes the dimension of \textbf{x} from \textit{N} to \textit{m}, as \textbf{x} transforms to \textbf{y}. So, the computational complexity of root-MUSIC algorithm changes from \(\mathcal{O}(N^3 + N^2M + \text{degree}-N~ \text{rooting})\) to \(\mathcal{O}(m^3 + m^2M + \text{degree}-m~ \text{rooting})\). Thus, it can be seen that CS beamformer greatly reduces computational complexity without affecting resolution of the algorithm 
\section{Subspace deviation in CS beamformer}
In this section, first the significance of study of subspace deviation is discussed and next, the same is derived for the proposed algorithm. \\
\textit{Significance of this analysis}: If the DOA being estimated by the algorithm comes out to be incorrect, the algorithm is said to have a performance breakdown. This performance breakdown for root-MUSIC algorithm has been studied in \cite{shag} in Step 1 analysis of the proposed algorithm by the authors. In the case of root-MUSIC algorithm, this breakdown will happen when the root of (8) is incorrect, or a root from the noise subspace (away from the unit circle in \textit{z}-domain) is selected. For very large number of snapshots, this occurs only in low SNR environment and inversely for low number of snapshots, in high SNR which is very undesirable for practical purposes. So to study the same for CS beamformer root-MUSIC algorithm, the following investigation is performed.
The estimated value of covariance matrix of \textbf{x} is given as $\widehat{\textbf{R}}_{xx}$ , where $\widehat{\textbf{R}}_{xx} = \frac{1}{T}\sum_{t=1}^T \textbf{xx}^H$ as in practical applications, only finite samples are available. Putting (3) in above, we have
\begin{align}
\widehat{\textbf{R}}_{xx} = \frac{1}{T}\sum_{t=1}^{T}(\textbf{A}\textbf{s}(t) + \textbf{w}(t))(\textbf{A}\textbf{s}(t) + \textbf{w}(t))^H\nonumber\\
= \textbf{A}\left\{\frac{1}{T}\sum_{t=1}^{T}\textbf{s}(t)\textbf{s}^H(t)\right\}\textbf{A}^H + \left\{\frac{1}{T}\sum_{t=1}^{T}\textbf{w}(t)\textbf{w}^H(t)\right\} \nonumber\\+ \underbrace{\textbf{A}\left\{\frac{1}{T}\sum_{t=1}^{T}\textbf{s}(t)\textbf{w}^H(t)\right\} + \left\{\frac{1}{T}\sum_{t=1}^{T}\textbf{w}(t)\textbf{s}^H(t)\right\}\textbf{A}^H}_\textbf{undesirable terms}
\end{align} 

In (9), the first two terms of $\widehat{\textbf{R}}_{xx}$ represent the signal and noise constituents, respectively. The last two terms in (9) are undesirable terms which are estimates for the correlation between the signal and noise components. As the noise vectors are assumed to be zero-mean and independent of the signal vectors, they are uncorrelated to each other. Hence, in case of large number of snapshots , these two last terms ideally tend  to zero. However, the number of available snapshots can be limited in practical applications. But in that case, these terms in (9) will contribute more to $\widehat{\textbf{R}}_{xx}$ and cause the signal and noise eigenvector (subspaces) estimates to deviate from their true subspaces. 

With the introduction to CS beamformer to reduce the data requirement, it becomes a necessity to analyse the effect of low snapshot number on the performance of the proposed CS beamformer root-MUSIC algorithm. As we are using the CS-DOA model (4) for the direction estimation, (9) becomes 
\begin{align}
\widehat{\textbf{R}}_{yy} = \textbf{B}\left\{\frac{1}{T}\sum_{t=1}^{T}\textbf{s}(t)\textbf{s}^H(t)\right\}\textbf{B}^H + \left\{\frac{1}{T}\sum_{t=1}^{T}\textbf{w}(t)\textbf{w}^H(t)\right\} \nonumber\\+ \underbrace{\textbf{B}\left\{\frac{1}{T}\sum_{t=1}^{T}\textbf{s}(t)\textbf{w}^H(t)\right\} + \left\{\frac{1}{T}\sum_{t=1}^{T}\textbf{w}(t)\textbf{s}^H(t)\right\}\textbf{B}^H}_\textbf{undesirable terms}
\end{align}
\noindent where $\widehat{\textbf{R}}_{yy} = \frac{1}{T}\sum_{t=T}^T\textbf{yy}^H$ and $\textbf{B} = \Phi \textbf{A}$.
Define \textbf{Q} $\triangleq [\textit{v}_{n1}, \textit{v}_{n2}, ..., \textit{v}_{n(m-M)}]$ and \textbf{P} $\triangleq [\textit{v}_{s1}, \textit{v}_{s2}, ..., \textit{v}_{s(M)}]$ where $\textit{v}_{ni}$ ($\textit{i}^{th}$ column of \textbf{Q}) and $\textit{v}_{sj}$ ($\textit{j}^{th}$ column of \textbf{P}) are the noise and signal eigenvectors respectively. In \cite{shag}, the subspace deviation is defined as the average value of the energy of the estimated signal eigenvectors deviating to the true noise subspace. Mathematically, it is defined as
\begin{equation}
\xi = \frac{1}{M}\sum_{j=1}^{M}\parallel\Gamma_N\textit{v}_{sj}\parallel_2^2
\end{equation}  
\noindent where \(\Gamma_N \triangleq \textbf{Q}\textbf{Q}^H\). 
The expression in (11), for the case of CS beamformer root-MUSIC algorithm gets simplified to 
\begin{equation}
\xi = \frac{1}{M}\text{Tr}\{\textbf{V}^\dagger \triangle \textbf{R}_{yy} \Gamma_N \triangle \textbf{R}_{yy} \textbf{V}^\dagger\}
\end{equation}  
\noindent where $\triangle \textbf{R}_{yy}= \textbf{R}_{yy} - \widehat{\textbf{R}}_{yy}$, \(\textbf{R}_{yy} =\textit{E}[\textbf{yy}^H]\) and $\textbf{V}^\dagger$ is the pseudo-inverse of $\textbf{V} = \textbf{R}_{yy} - \sigma_n^2\textbf{I}_m$. Its expected value is given by 
\begin{equation}
E[\xi] = \frac{\sigma_n^2}{TM}\sum_{j=1}^{M}\frac{\lambda_{j+1}}{( \lambda_{j+1} - \sigma_n^2 )^2}
\end{equation}  

\noindent where $\lambda_j$ represents the eigenvalue of $\textbf{R}_{yy}$. For CS beamformer root-MUSIC, $\xi$ in (12) is derived same as expressed in (26) of \cite{shag} as $\rho$ with \textbf{V} defined as above and $\triangle \textbf{R} = \triangle \textbf{R}_{yy}$. The derivation of $\textit{E}[\xi]$ is given in Appendix A. The equation (13) is exclusive to the proposed CS beamformer root-MUSIC algorithm and its comparison with respect to (27) in \cite{shag} is discussed in the following two remarks. \\
\textit{Remark} 1: In (27) of \cite{shag}, the expected value of the subspace deviation is proportional to $\frac{\sigma_n^2}{T}$ (number of snapshot in \cite{shag} is denoted by \textit{N}, for clarity we continue our notation). In (13) of this paper, the same is true . This shows that deviation of signal subspace is significant for a small number of snapshot as well as low SNR value.\\
\textit{Remark} 2: The significant difference in (13) and (27) of \cite{shag} is that (27) depends on number of sensor whereas (13) is independent of this. This is because in (13), \textit{N} - \textit{M} = 1, hence making value of $\textit{E}[\xi]$ comparatively farther from $\xi$ than $\textit{E}[\rho]$ to $\rho$, confirming the expected behaviour of the proposed algorithm in comparison to the root-MUSIC algorithm (See fig. 3).  

In the next section, MATLAB simulations are being demonstrated with analysis to validate the proposed work. For clarity, it is pointed out that theoretical subspace deviation is given by (12) and the simulated subspace deviation is given by (13).    
 
\section{Simulation results and Analysis}
\noindent \textit{Example} \textbf{1} : The DOA estimation in this example is performed in the following scenario. The SNR is kept as 15 dB. The ULA is chosen with \textit{N} = 7 sensors. Two non-coherent sources are considered with directions 20\degree , -50\degree). The measurement matrix, $\Phi$ is chosen to be random Gaussian matrix with dimensions $\textit{m}\times \textit{N}$ where \textit{m} = 8. Fig.1 shows the plot for this simulation in which \textit{T} = 1000. 
\begin{figure}[h]
   \centering
    \includegraphics[height=5.5cm]{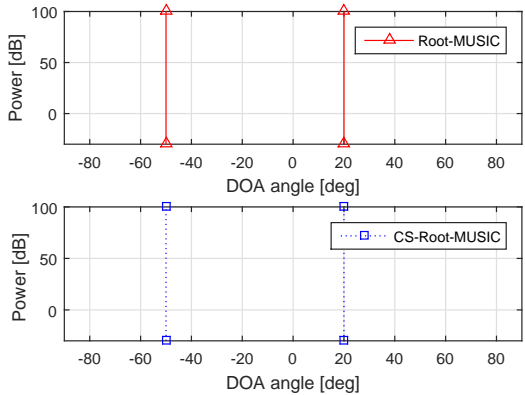}     
    \tiny\caption{ DOA estimation}
\end{figure}
\begin{figure}
   \centering
    \includegraphics[height=6.0cm]{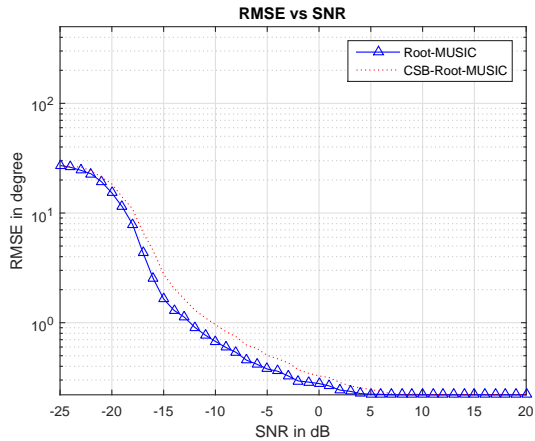}
    \caption{RMSE \textit{vs} SNR}
\end{figure}

From fig.1, it is observed that the CS beamformer is able to detect all the DOAs correctly, with same resolution as root-MUSIC algorithm. The difference between the two algorithms, is the effective estimated covariance matrix and the effective sensor array. In case of root-MUSIC it is $\widehat{\textbf{R}} = \widehat{\textbf{R}}_{xx} \in \displaystyle\mathbb{C}^{\textit{N$\times$N}}$, where as for CS beamformer root-MUSIC, it is $\widehat{\textbf{R}} = \widehat{\textbf{R}}_{yy} \in \displaystyle\mathbb{C}^{\textit{m$\times$m}}$. Though the size of $\widehat{\textbf{R}}$ is reduced drastically, the performance remains highly accurate in results. The sensor array of the root-MUSIC algorithm is same as the physical system, where as for the proposed algorithm, it changes to a virtual non- uniform sensor array due to the randomness on the measurement matrix $\Phi$. From fig.2, it can be concluded that with negligible deviation in root mean square error (RMSE) at different SNR, the performance of CS beamformer root-MUSIC algorithm is comparable with the root-MUSIC algorithm.\\*

\noindent \textit{Example} \textbf{2} : In this example, the subspace deviation \textit{vs} SNR for different snapshots is evaluated with the theoretical approximation of (13). First it is done for root-MUSIC algorithm in comparison CS beamformer root-MUSIC algorithm with \textit{T} = 10 , and then for different snapshot scenarios in the case of the proposed algorithm. The ULA is chosen with \textit{N} = 10 sensors. The number of sources is \textit{M} = 2.
\begin{figure}[h]
   \centering
    \includegraphics[height=6.0cm]{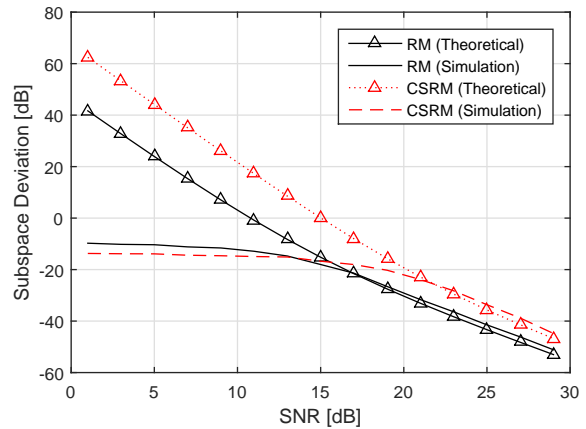}
    \caption{Subspace deviation comparison [root-MUSIC (RM), CS beamformer root-MUSIC (CSRM)]}
\end{figure} 
\begin{figure}
   \centering
    \includegraphics[height=5.0cm]{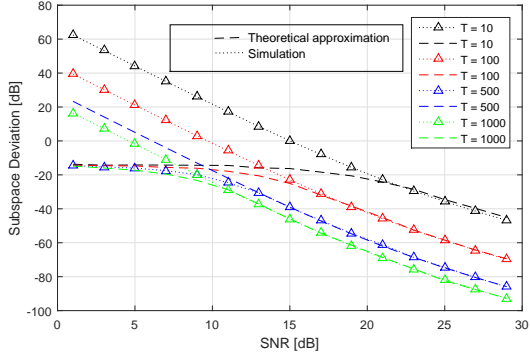}
    \caption{Subspace deviation for different \textit{T}'s}
\end{figure}

From fig.3, it is observed that the proposed CS beamformer root-MUSIC algorithm is having the breakdown point just ahead of the root-MUSIC algorithm when the number of snapshots is 10. The different snapshot scenarios of subspace deviation is investigated for the proposed algorithm (fig.4). It can be observed that as the number of snapshots increases, the breakdown moves towards lower SNR region, which is desirable.

\section{Conclusion}
In this paper, a novel CS beamformer root- MUSIC algorithm is proposed with a strict bound on the measurement matrix. The proposed algorithm not only maintains the resolution of root-MUSIC algorithm, but also performs comparatively with root-MUSIC algorithm under low snapshot scenario. Also, the subspace deviation in case of the proposed algorithm is discussed and analysed for different snapshot numbers. The proposed algorithm for all its restrictions on data model, displays robustness in adverse scenarios, and also provides an option of non-uniform arrangement of sensors though in linear manner.






\begin{thebibliography}{6}

\bibitem{naidu}
Naidu, P.S., "Sensor Array Signal Processing",\emph{1st ed. CRC Press; Boca Raton, FL, USA}, \hskip 1em plus
  0.5em minus 0.4em\relax 2001.

\bibitem{candes}
Candès, E.J., Wakin, M.B., "An Introduction To Compressive Sampling",\emph{in IEEE Signal Processing Magazine, vol.25, no.2}, \hskip 1em plus
  0.5em minus 0.4em\relax pp.21-30, March 2008.
    
\bibitem{ying}
Ying Wang, Geert Leus, Ashish Pandharipande, " Direction Estimation Using Compressive Sampling Array Processing",\emph{in IEEE/SP 15th workshop on SSP}, \hskip 1em plus
  0.5em minus 0.4em\relax pp 626-629, 2009.


\bibitem{schmidt}
Schmidt, R.O., "Multiple Emitter Location and Signal Parameter Estimation",\emph{IEEE Trans. On AP, vol.Ap-34, no.3}, \hskip 1em plus
  0.5em minus 0.4em\relax  pp. 276-288, 1986.
  
\bibitem{shag}
Shaghaghi M. and Vorobyov S. A., "Subspace Leakage Analysis and Improved DOA Estimation With Small Sample Size," \emph{IEEE Transactions on Signal Processing, vol. 63, no. 12}, \hskip 1em plus
  0.5em minus 0.4em\relax pp. 3251-3265, June15, 2015.
  
\bibitem{aich}
Aich A. and Palanisamy P., "A strict bound for dimension of measurement matrix for CS beamformer MUSIC algorithm," \emph{2016 IEEE Region 10 Conference (TENCON) Singapore, 2016}, \hskip 1em plus
  0.5em minus 0.4em\relax pp. 2602-2605.

\bibitem{bara}
Barabell A., "Improving the resolution performance of eigenstructure-based direction-finding algorithms," \emph{ICASSP '83. IEEE International Conference on Acoustics, Speech, and Signal Processing, 1983}, \hskip 1em plus
  0.5em minus 0.4em\relax pp. 336-339.
  
\bibitem{krim}
Krim H., Forster P., and Proakis J. G., "Operator approach to performance analysis of root-MUSIC and root-min-norm," \emph{IEEE Trans. Signal Process., vol. 40, no. 7}, \hskip 1em plus
  0.5em minus 0.4em\relax pp. 1687–1696, Jul. 1992.

\end{thebibliography}
%

\appendices 
\section{Derivation of (13)}
So continuing from (12), we use following two Lemmas in the case of CS-DOA model, whose proof can be directly extended from \cite{krim}.
\begin{lemma}
For all arbitrary matrices $\normalfont\textbf{C}_1$, $\normalfont\textbf{C}_2$	$\in \displaystyle\mathbb{C}^{\textit{k$\times$k}}$, we have
\begin{align}
E[\triangle \normalfont \textbf{R}\textbf{C} \triangle \textbf{R}] = \frac{1}{T}\text{Tr}\{\normalfont \textbf{R}\textbf{C}_1\}\textbf{R}
\end{align}
and
\begin{align}
E[\normalfont \text{Tr}\{ \triangle \textbf{R}\textbf{C}_1\}\text{Tr}\{\triangle \textbf{R} \textbf{C}_2\}] = \frac{1}{T}\text{Tr}\{\normalfont \textbf{R}\textbf{C}_1\textbf{R}\textbf{C}_2\}
\end{align}  
\end{lemma}
Using (14) in (12), the expected value of $\xi$ is given as 
\begin{align}
E[\xi] = \frac{1}{M}\text{Tr}\{\textbf{V}^\dagger E[\triangle \textbf{R}_{yy} \Gamma_N \triangle \textbf{R}_{yy}] \textbf{V}^\dagger\}\nonumber \\
= \frac{1}{M}\text{Tr}\{\textbf{V}^\dagger \frac{1}{T}\text{Tr}\{\textbf{R}_{yy} \Gamma_N\}\textbf{R}_{yy} \textbf{V}^\dagger\}\nonumber \\
= \frac{1}{MT}\text{Tr}\{\textbf{R}_{yy} \Gamma_N\} \text{Tr}\{\textbf{V}^\dagger\textbf{V}^\dagger\textbf{R}_{yy}\}
\end{align}  
Using the fact that steering vectors are orthogonal to the noise subspace, we have
\begin{equation}
\text{Tr}\{ \Gamma_N\textbf{R}_{yy}\} = \sigma_n^2
\end{equation}
and
\begin{equation}
\textbf{V}^\dagger\textbf{V}^\dagger\textbf{R}_{yy} = \sum_{j=1}^{M}\frac{\lambda_{j+1}}{(\lambda_{j+1} - \sigma_n^2)^2}\textbf{v}_{sj}\textbf{v}_{sj}^H
\end{equation}
This implies 
\begin{equation}
\text{Tr}\{ \textbf{V}^\dagger\textbf{V}^\dagger\textbf{R}_{yy}\} = \sum_{j=1}^{M}\frac{\lambda_{j+1}}{(\lambda_{j+1} - \sigma_n^2)^2}
\end{equation}
Putting (19) and (17) in (16), gives (13) and completes the derivation.

\end{document}